\documentclass[aps,prl,twocolumn,floats,amsmath,amssymb,showpacs]{revtex4}

\usepackage{graphicx}
\usepackage{dcolumn}
\usepackage{psfrag}

\begin{document}
\title{Caustic activation of rain showers}
\author{Michael Wilkinson$^{1}$, Bernhard Mehlig$^{2}$ and Vlad Bezuglyy$^{2}$}
\affiliation{ $^{1}$Faculty of Mathematics and Computing, The Open
University, Walton Hall,
Milton Keynes, MK7 6AA, England \\
$^{2}$Department of Physics, G\"oteborg University, 41296
Gothenburg, Sweden \\}

\begin{abstract}
We show quantitatively how the collision rate of droplets of
visible moisture in turbulent air increases very abruptly as the
intensity of the turbulence passes a threshold, due to the
formation of fold caustics in their velocity field. The formation
of caustics is an activated process, in which a measure of the
intensity of the turbulence, termed the Stokes number ${\rm St}$,
is analogous to temperature in a chemical reaction: the rate of
collision contains a factor $\exp(-C/{\rm St})$. Our results are
relevant to the long-standing problem of explaining the rapid
onset of rainfall from convecting clouds. Our theory does not
involve spatial clustering of particles.
\end{abstract}


%PACS:
%45.50 Tn Collisions
%47.55 D Drops and bubbles
%92.60 Mt Particles and aerosols
%92.60 Nv Cloud physicsb

\maketitle

\begin{figure}[b]
\includegraphics[width=6cm,clip]{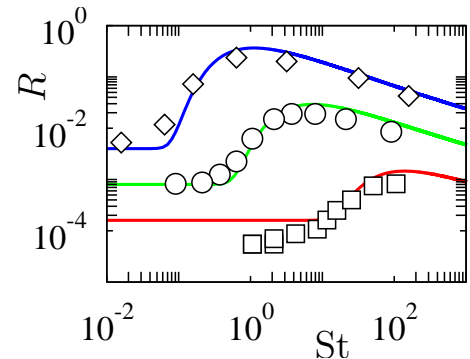}
\caption{\label{fig: 1}
Particle collision rate as a function of Stokes number, for three
different values of the Kubo number (defined in equation (\ref{eq:
1.3})). In these simulations we used $n=10^{3}$, $a=2.5\times
10^{-4}$, $\eta=0.1/\pi$, $\tau=0.1$, and varied $u_0$ and $\gamma
$: ${\rm Ku}=1$ ($\Diamond$), ${\rm Ku}=0.2$ ($\circ$), and ${\rm
Ku}=0.04$ ($\Box$). The theoretical curves are (\ref{eq: 1.7}),
with actions $S$ obtained from figure \ref{fig: 3}. The constant
$C_{\rm a}$ in $R_{\rm a} = C_{\rm a} na^d u_0/\eta$ was fitted
(see text). }
\end{figure}

It is common experience that rainfall can commence very abruptly
from cumulus clouds (which form when the atmosphere is convecting)
but has a much slower onset from stratiform clouds in a stable
atmosphere. This can happen even when no part of the cloud is
below freezing point. It is believed that the difference between
convecting and stable clouds arises because the convection gives
rise to small-scale turbulent motion which facilitates the
coalescence of microscopic water droplets (\lq visible moisture')
into raindrops. This idea has a long history (\cite{Saf56} is a
significant contribution containing references to other early
papers), but a satisfying theory has been elusive and the topic
remains a subject of intensive research, reviewed recently in
\cite{Sha03}. Numerical experiments have shown a dramatic increase
in the rate of collision of suspended particles when the intensity
of turbulence exceeds a certain threshold. This effect was first
described in \cite{Sun97}, see also \cite{Zho98}. Here we present
a simple quantitative theory of this phenomenon, illustrated in
figure \ref{fig: 1}, which shows the collision rate $R$ of an
aerosol as a function of a dimensionless parameter termed the
Stokes number, ${\rm St}$, which contains information about the
turbulence intensity, $\epsilon$, and the radius of the water
droplets, $a$ (other symbols in the caption are defined later).
There is a precipitous increase in the collision rate at a
threshold value of ${\rm St}$, which was also observed in \cite
{Sun97,Zho98}. The current consensus, represented in
\cite{Sha03,Sun97,Zho98,Max87}, is that the increased rate of
collision involves spatial clustering of particles. One exception
is \cite{Fal02}, which presents a theory having elements
(described later) in common with our own, but which is more
complex in its formulation and less precise in its conclusions.
Spatial clustering plays no role in our theory.

The properties of clouds are very variable and the sizes of
visible moisture droplets have a large dispersion, but typically
the average of the radius is approximately $10\mu {\rm m}$ and the
density is $n\approx 10^8{\rm m}^{-3}$ \cite{Sha03}. The motion of
the droplets is dominated by viscous forces, so that the equation
for the position $\mbox{\boldmath$r$}$ of a droplet is well
approximated by
\begin{equation}
\label{eq: 1.1} \ddot{\mbox{\boldmath$r$}}=\gamma
(\mbox{\boldmath$u$}(\mbox{\boldmath$r$},t)-\dot{\mbox{\boldmath$r$}})
\end{equation}
until the particles come into contact \cite {Saf56} (here
$\mbox{\boldmath$u$}(\mbox{\boldmath$r$},t)$ is the velocity field
of the air and dots denote derivatives with respect to time).
According to Stokes's formula for the viscous drag on a sphere,
$\gamma=9\rho_{\rm g}\nu/2\rho_{\rm f}a^2$ ($\nu$ being the
kinematic viscosity of air, $\rho_{\rm f}$ and $\rho_{\rm g}$ the
densities of water and air respectively): using the values above
we estimate $\gamma\approx 500\,{\rm s}^{-1}$. The turbulent
motion is a multi-scale flow with an approximately power-law
spectrum of spatial fluctuations of
$\mbox{\boldmath$u$}(\mbox{\boldmath$r$},t)$ \cite{Fri97}.
According to the Kolmogorov theory of turbulence, the short
wavelength cutoff of the power-law spectrum, $\eta$ (termed the
Kolmogorov length) is a function of $\nu $ and of the rate of
dissipation per unit mass in the turbulence, $\epsilon$. Similar
considerations apply to the shortest characteristic timescale,
$\tau$. Dimensional arguments \cite{Fri97} then imply
\begin{equation}
\label{eq: 1.2} \eta \sim
({\nu^3/{\epsilon}})^{1/4} \ ,\ \ \ \tau \sim
({\nu/ {\epsilon}})^{1/2}\ .
\end{equation}
The typical velocity difference for points separated by the
correlation length $\eta$ will be denoted by $u_0$. The
dimensional arguments of the Kolmogorov theory imply that $u_0\sim
\eta/\tau$ for turbulent flow. The dissipation rate $\epsilon$ is
highly variable between different clouds. Values of
$\epsilon=0.1{\rm m}^3{\rm s}^{-2}$ are typical \cite{Sha03};
$\nu\approx 10^{-5}{\rm m}^2{\rm s}^{-1}$ for air in the lower
atmosphere, giving $\eta\approx 3\times 10^{-4}{\rm m}$, $\tau
\approx 10^{-2}{\rm s}$.

We can form four independent dimensionless parameters describing
the microscopic motion of the water droplets: these are
\begin{equation}
\label{eq: 1.3} {\rm St}={1\over{\gamma \tau}}\ ,\ \ \ {\rm
Ku}={u_0\tau\over{\eta}}\ , \ \ \ n\eta^d\ ,\ \ \ \eta/a
\end{equation}
($d$ is the dimensionality of space). The values quoted above give
${\rm St}\approx 0.2$. The Kubo number ${\rm Ku}$ cannot be large
\cite{Dun05} and the comments above indicate that it is of order
unity for fully developed turbulence, but small values of ${\rm
Ku}$ occur in stirred fluids, and this parameter will play a role
later. From the data above we see that $n\eta^d\ll 1$ and $\eta/a
\gg 1$ for water droplets in clouds.

Saffman and Turner \cite{Saf56} only considered the case of ${\rm
St}\ll 1$ in detail, because they believed that ${\rm St}\gg 1$
would only be realised in the most unstable cumulonimbus clouds.
Later it was understood that fully-developed turbulent motion with
high Reynolds number can exhibit pronounced intermittency
\cite{Fri97}. One consequence is that the rate of dissipation can
fluctuate wildly relative to the mean value, and measurements in
clouds have indicated that $\epsilon$ can have an approximately
log-normal distribution \cite{Sha03}. The notion that
intermittency can promote the formation of rain showers by
creating localised regions of very high turbulent intensity is
considered in \cite{Sha03} and \cite{Fal02}.

Aerosol particles in a turbulent flow may exhibit a degree of
clustering: mechanisms for this process were proposed in
\cite{Max87,Som93,Fal02} and reference \cite{Dun05} describes
recent advances. Most of the recent literature on the initiation
of rainfall by turbulence assumes that the mechanism involves
clustering \cite{Sha03,Sun97,Zho98,Max87}. This theory is
unsatisfactory for a number of reasons. Firstly, the widely
accepted view is that clustering is due to particles being
centrifuged away from regions of high vorticity. It is argued that
this effect is strongest when ${\rm St}$ is close to unity: it
does not occur when the particle motion is highly damped (${\rm
St}\ll 1$), or when the vortices are too short-lived (${\rm St}\gg
1$). Figure \ref{fig: 1} shows, however, that the collision rate
rises abruptly at a threshold value of ${\rm St}$ (which depends
upon ${\rm Ku}$) and remains high for ${\rm St}\gg 1$: this is
equally true for simulations of single-scale flows, such as figure
\ref{fig: 1}, and multi-scale flows, such as those in
\cite{Sun97}. Another difficulty is that the clustering can only
occur for particle separations smaller than $\eta$, but in a cloud
the density of particles is so low that there is unlikely to be
more than one particle in a cube of length $\eta$. Also, it has
not been established that the centrifuge mechanism is sufficiently
effective to explain the large increase in collision rate.

\begin{figure}[b]
\includegraphics[width=5.5cm,clip]{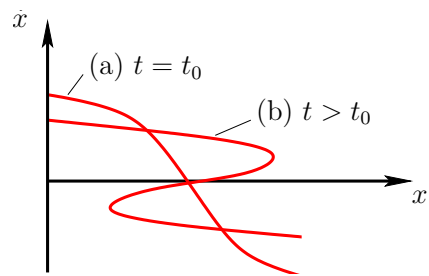}
\caption{\label{fig: 2}
Schematic showing the formation of a caustic: (a) Initial
configuration, showing manifold representing droplet velocity as a
function of position. (b) This manifold has developed fold caustics.
The velocity field is multivalued between the folds.}
\end{figure}

In our theory, the dramatic increase in the collision rate is a
consequence of the formation of \lq caustics', illustrated
schematically in figure \ref{fig: 2}. Here the velocity of the
water droplets as a function of position is initially
single-valued  (curve (a)) but droplets with a large velocity
overtake slower-moving particles, so that at a later time the
droplet velocity is multi-valued (curve (b)). The region where the
velocity is a multi-valued function is bounded by two fold
caustics. The caustics have two effects which could enhance the
collision rate of water droplets. Firstly, at the caustic lines
themselves there can be a divergence in the density of particles.
This effect is discussed in \cite{Fal02,Wil05}: it is analogous to
the divergence of light intensity on optical caustics
\cite{Ber81}, but we argue that it is not relevant to collisions
between microscopic water droplets in clouds, where $n\eta^3\ll
1$. The other effect of the caustics is that when the velocity
field is multi-valued, droplets at the same position are moving
with differing velocity, and their relative motion produces
collisions. This has no analogue in optical caustics. The
importance of this effect was previously emphasised in
\cite{Fal02}.

Our theory must consider the rate of collision between droplets
both with and without caustics, and the rate at which caustics are
formed. Saffman and Turner \cite{Saf56} discussed the collision
(due to shearing motion) of droplets which are advected with the
air: this approximation is valid in the limit as ${\rm St}\to 0$;
we term this advective collision rate $R_{\rm a}$. At large Stokes
numbers a theory due to Abrahamson \cite{Abr75} treats the water
droplets as a gas of particles with random and uncorrelated motion
and calculates the collision rate $R_{\rm g}$ by gas-kinetic
theory. This theory is widely regarded as being applicable only at
very large Stokes numbers, and indeed the two theories give
results which differ by a factor of order $\eta/a$ at ${\rm
St}\approx 1$. This might be taken as evidence that at least one
of these theories does not work when the Stokes number is of order
unity. However, we shall argue that the gas kinetic model is
applicable as soon as caustics have formed and that the collision
rate is well-approximated by $R=R_{\rm a}+f({\rm St},{\rm
Ku})R_{\rm g}$, where $f({\rm St},{\rm Ku})$ is the fraction of
the coordinate space for which the velocity field has become
multi-valued due to the formation of caustics.

\begin{figure}[b]
\includegraphics[width=6cm,clip]{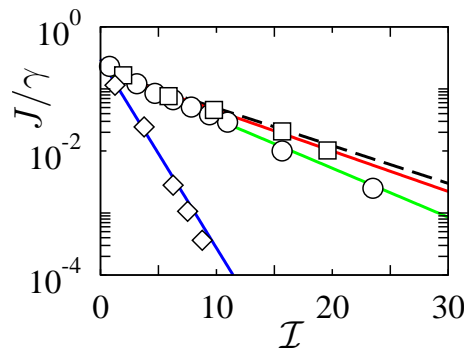}
\caption{\label{fig: 3}
Rate of caustic formation as a function of ${\cal I}\propto {\rm
Ku}^2 {\rm St}$, showing evidence for \lq activated' behaviour:
${\rm Ku}=1$ ($\Diamond$), with action $S=0.70$, ${\rm Ku}=0.2$,
$S=0.18$ ($\circ$), and ${\rm Ku}=0.04$, $S=0.15$ ($\Box$). Also
shown is the limiting behaviour as Ku $\rightarrow 0$, with action
$S=0.14$ (dashed line).}
\end{figure}

First we consider the rate of production of caustics in more
depth. In earlier works (\cite{Wil03,Wil05}), two of the present
authors described  the formation of caustics in the limit where
${\rm Ku}\ll 1$. In the one-dimensional case, we considered the
linearisation of equation (\ref{eq: 1.1}), describing the small
separation in space $\delta x$ and velocity $\delta v$ for two
nearby droplets. The equation of motion for $X=\delta v/\delta x$
is \cite{Pit01}
\begin{equation}
\label{eq: 1.4} {{\rm d}X\over{{\rm d}t}}=-\gamma X-X^2+{\partial
u\over{\partial x}}(x(t),t)
\end{equation}
which is a stochastic differential equation, in which the velocity
gradient acts as a random forcing term. The droplets encounter a
caustic whenever $\delta x$ passes through zero, implying that $X$
goes to infinity in one direction, then jumps instantaneously to
the reflected point at infinity. When fluctuations drive $X$ to a
sufficiently negative value, it will almost certainly escape to
$-\infty$ in a finite time. In \cite{Wil03}, equation (\ref{eq:
1.4}) was considered in the limit ${\rm St}\to \infty$, ${\rm
Ku}\to 0$, where the dynamics of $X$ can be approximated by a
Langevin process. The associated Fokker-Planck equation was solved
exactly, giving  the rate $J$ at which any one droplet passes
through caustics. Similar results were obtained in two
\cite{Wil05} and three \cite{Dun05} spatial dimensions. They are
expressed in terms of a dimensionless parameter ${\cal I}$ defined
in terms of the strain-rate correlation function
\begin{equation}
\label{eq: 1.5} {\cal I}={1\over{2\gamma}} \int_{-\infty}^\infty
{\rm d}t\ \biggl\langle {\partial u_1\over{\partial
x_1}}(\mbox{\boldmath$r$}(t),t){\partial u_1\over{\partial
x_1}}(\mbox{\boldmath$r$}(0),0)\biggr\rangle
\end{equation}
(we use angular brackets to indicate averages). In the limit as
${\cal I}\to 0$, we find that the rate of caustic formation is
determined by the rate of escape due to diffusion of the
coordinate $X$ from an attractive fixed point at $X=0$. The escape
rate is asymptotic to
\begin{equation}
\label{eq: 1.6} J=J_0 \exp(-S/{\cal I})
\end{equation}
(for some constant $J_0$), where in $S$ can be obtained as the
action of a trajectory of a Hamiltonian function
\cite{Meh04,Fre84}. Note that we can write ${\cal I}\sim {\rm
Ku}^2{\rm St}$, so that (\ref{eq: 1.6}) is consistent with the
alternative form quoted in the abstract. In one dimension we
extract $S=1/6$ from our exact expression for $J$, obtained in
\cite{Wil03}. In two and three dimensions, one obtains $S\approx
0.14$ \cite{Wil05} and $S\approx 0.12$ \cite{Dun05} in the limit
as ${\rm Ku}\to 0$. Our numerical results, summarised in figure
\ref{fig: 3}, show that the activated escape model for the
formation of caustics, described by (\ref{eq: 1.6}), is valid even
when ${\rm Ku}$ approaches unity. We find that the action $S$
depends upon ${\rm Ku}$. For our model (described below) we found
$S=0.70$ for ${\rm Ku}=1$, $S=0.18$ for ${\rm Ku}=0.2$ and
$S=0.15$ for ${\rm Ku}=0.04$, consistent with the limiting value
$S=0.14$ for ${\rm Ku}\to 0$ quoted above.

The dependence of $J$ upon ${\cal I}$ (or equivalently
, on ${\rm St}$) is analogous to the dependence of the rate of a
chemical reaction on temperature $T$, which is well approximated
by expressions containing an Arrhenius factor, $\exp(-E/kT)$ ($E$
is the activation energy, $k$ is the Boltzmann constant). In this
sense we may regard the formation of caustics as an activated
process, in which the dimensionless intensity of the turbulence
(${\cal I}$ or ${\rm St}$) plays a role analogous to temperature.

We argue that the function $f({\rm St},{\rm Ku})$ describing the
prevalence of caustics is closely related to the rate of caustic
formation, so that the collision rate is well approximated by
\begin{equation}
\label{eq: 1.7} R=R_{\rm a}+\exp(-S/{\cal I})R_{\rm g}\ .
\end{equation}
Formulae for the advective and gas-kinetic collision rates,
$R_{\rm a}$ and $R_{\rm g}$ respectively, are given below. It is
equation (\ref{eq: 1.7}) which is the theoretical curve
illustrated in figure \ref{fig: 1}, using values of $S$ obtained
from figure \ref{fig: 3}. We fitted the prefactor of the
expression for $R_{\rm a}$, which is the dominant term when ${\rm
St}\ll 1$, but there are no other fitted parameters. We remark
that combining equations (2) and (4) of \cite{Fal02} gives an
expression which can be written in the form $R=R_{\rm a}+P\,R_{\rm
c}$, analogous to our (\ref{eq: 1.7}), in which $R_{\rm c}$ is
similar to the precise asymptote, $R_{\rm g}$. The expression for
the factor $P$ in \cite{Fal02} has a different dependence upon
${\rm St}$.

Our numerical simulations use a two-dimensional model in which the
divergenceless velocity
$\mbox{\boldmath$u$}(\mbox{\boldmath$x$},t)$ is obtained from a
scalar random field $\psi(\mbox{\boldmath$x$},t)$ by writing
$\mbox{\boldmath$u$}=(\partial \psi/\partial y,-\partial
\psi/\partial x)$. The field $\psi(\mbox{\boldmath$x$},t)$ is a
Gaussian random function with spatially and temporally stationary,
isotropic statistics. The mean value is zero and the correlation
function is
$\langle\psi(\mbox{\boldmath$x$},t)\psi(\mbox{\boldmath$x$}',t')\rangle=
u_0^2 \exp[-(t-t')^2/2\tau^2]\exp(-\vert
\mbox{\boldmath$x$}-\mbox{\boldmath$x$}'\vert^2/2\eta^2)$, which
gives ${\cal I}=\sqrt{\pi/2}\,{\rm Ku}^2{\rm St}$ for this model.
The droplets are initially randomly positioned, and they are
regarded as having collided when their separation falls below $2a$
(that is, we assume that the collision efficiency is unity
\cite{Saf56}). Figure \ref{fig: 1} shows the rate of collision $R$
for a single particle. We did not include gravitational settling
or other effects which also occur without turbulence. The results
for our single-scale flow are very similar to the simulations
using Navier-Stokes flows, reported in \cite{Sun97,Zho98},
implying that the multi-scale aspect of those simulations is not
essential to their understanding.

In order to complete the discussion of our formula (\ref{eq: 1.7})
for the collision rate we describe the asymptotic formulae for the
advective and gas-kinetic collision rates. In the advective limit
the suspended particles are brought into contact by the effect of
shearing motion in the flow. The typical shear rate is $u_0/\eta$.
The relative velocity for particles that will be brought into
contact is $\Delta v\sim a u_0/\eta$, and the volume swept in time
$\Delta t$ is $\Delta V\sim a^du_0\Delta t/\eta$. Collisions
typically occur when $n\Delta V\sim 1$, so that the expected rate
of collision is $R_{\rm a}=C_{\rm a} na^du_0/\eta$ (where $C_{\rm
a}$ is a constant which we fitted numerically). Saffman and Turner
\cite{Saf56} gave an upper bound on the collision rate which can
be calculated analytically if correlation functions of the flow
field are known. Their approach gives a precise asymptote for the
steady hyperbolic flow which they discussed, but in general it is
an upper bound because their calculation does not account for
multiple collisions in generic flows (which are not steady and
which may be locally elliptic).

When ${\rm St}\gg 1$, the inertia of the aerosol particles means
that their motion becomes uncorrelated with the velocity of fluid,
and a gas-kinetic model of the type proposed by Abrahamson
\cite{Abr75} is appropriate. The maximal Lyapunov exponent
$\lambda$ for motion of the particles is positive \cite{Meh04} and
if the rate of collision satisfies $R\ll \lambda$, then the
velocities of nearby particles are completely randomised by the
time they collide (our estimates below verify this assumption).
The aerosol particles become a gas of droplets with velocities
which are uncorrelated with each other and with that of the
turbulent air. The velocity distribution is Maxwellian, even if
$\mbox{\boldmath$u$}$ is not Gaussian distributed, because the
equation of motion for the droplet velocity is analogous to an
Ornstein-Uhlenbeck process. The rate of collision is exactly the
same as for a hard-sphere gas with the same particle radius and
r.m.s. velocity. The number of collisions per unit time of a given
particle in a dilute gas is $R_{\rm g}\sim na^{d-1}\bar v_{\rm r}$
($R_{\rm g}=4an\bar v_{\rm r}$ in two dimensions), where $\bar
v_{\rm r}$ is the mean relative speed of the droplets. For a
Maxwellian velocity distribution, $\bar v_{\rm r}=\sqrt{\pi/
4}\sqrt{\langle \mbox{\boldmath$v$}_{\rm r}^2\rangle}$ in two
dimensions and we have $R_{\rm g}=2\sqrt{\pi}na\sqrt{\langle
\mbox{\boldmath$v$}_{\rm r}^2\rangle}$. It remains to calculate
the r.m.s. velocity of the water droplets, $\langle
\mbox{\boldmath$v$}^2\rangle$, from which we obtain $\langle
\mbox{\boldmath$v$}^2_{\rm r}\rangle=2\langle
\mbox{\boldmath$v$}^2\rangle$. The solution of the equation of
motion (\ref{eq: 1.1}) for the velocity of a droplet is
\begin{equation}
\label{eq: 1.11} \mbox{\boldmath$v$}(t)=\gamma \int_{-\infty}^t
{\rm d}t'\
\exp[-\gamma(t-t')]\mbox{\boldmath$u$}(\mbox{\boldmath$r$}(t'),t')
\ .
\end{equation}
The variance of the velocity is obtained by squaring this
expression and averaging: when ${\rm St}\gg 1$ this is asymptotic
to
\begin{equation}
\label{eq: 1.12} \langle \mbox{\boldmath$v$}^2\rangle =
{\gamma\over {2}} \int_{-\infty}^\infty {\rm d}t\ \langle
\mbox{\boldmath$u$}({\bf 0},t)\cdot \mbox{\boldmath$u$}({\bf
0},0)\rangle \ .
\end{equation}
This gives a collision rate $R_{\rm g}\sim na^{d-1}
u_0\sqrt{\gamma \tau}$, (for our two-dimensional model, $R_{\rm
g}=2^{7/4}\pi^{3/4}nau_0\sqrt{\gamma \tau}$). Thus $R_{\rm
g}/R_{\rm a}$ is of order $\eta/a$ when ${\rm St}\approx 1$ and
${\rm Ku}\approx 1$.

In summary, we have shown that the collision rate is well
approximated by the gas-kinetic model as soon as the turbulence
intensity ${\cal I}$ exceeds the action $S$. Finally, we must
consider whether this gives a sufficiently rapid rate of
collision. In three dimensions, the gas-kinetic model gives a
collision rate $R_{\rm g}=16 \sqrt{\pi/3}na^2 \sqrt{\langle
\mbox{\boldmath$v$}^2\rangle}$. From (\ref{eq: 1.12}), we estimate
$\langle \mbox{\boldmath$v$}^2\rangle \sim \sqrt{{\rm
St}}\,\eta/\tau $. Using the above data, this estimate gives a
collision rate of $R_{\rm g}\approx 5\times 10^{-3}{\rm s}^{-1}$
when ${\rm St}\approx 1$. We conclude that rainfall can be
initiated in a timescale of a few minutes, provided that a
sufficiently large part of the cloud has a turbulence intensity,
${\cal I}$ which exceeds the action for forming caustics, $S$.

B. M. acknowledges support by Vetenskapsr\aa{}det.

\end{document}